\documentclass[twocolumn,english]{revtex4}
\usepackage[T1]{fontenc}
\usepackage[latin1]{inputenc}
\usepackage{amsmath}
\usepackage{graphicx}
\usepackage{amssymb}
\usepackage{graphicx}
\usepackage{graphicx}
\usepackage{babel}

\makeatletter
\providecommand{\LyX}{L\kern-.1667em\lower.25em\hbox{Y}\kern-.125emX\@}
\providecommand{\LyX}{L\kern-.1667em\lower.25em\hbox{Y}\kern-.125emX\@}

\def\be{\begin{equation}}
\def\ee{\end{equation}}
\def\bea{\begin{eqnarray}}
\def\eea{\end{eqnarray}}

\makeatother

\begin{document}

\preprint{astro-ph/05mmddd}
\title{Observational Constraints on Silent Quartessence}
\author{Luca Amendola}
\email{amendola@mporzio.astro.it}
\affiliation{INAF/Osservatorio Astronomico di Roma, Via Frascati 33, I-00040 Monte Porzio
Catone, RM, -- Italy}
\author{Ioav Waga}
\email{ioav@if.ufrj.br} \affiliation{Universidade Federal do Rio de
Janeiro, Instituto de F\'{i}sica, CEP 21941-972, Rio de Janeiro, RJ,
Brazil}
\author{Fabio Finelli}
\email{finelli@bo.iasf.cnr.it}
\affiliation{INAF/IASF, Sezione di Bologna, Istituto di Astrofisica
Spaziale e Fisica Cosmica, Istituto Nazionale di Astrofisica, via Gobetti,
101 -- I-40129 Bologna -- Italy}
\date{\today{}}

\begin{abstract}
We derive new constraints set by SNIa experiments (`gold' data sample of
Riess \emph{et al.}), X-ray galaxy cluster data (Allen \emph{et al.} \emph{%
Chandra} measurements of the X-ray gas mass fraction in 26
clusters), large scale structure (Sloan Digital Sky Survey spectrum)
and cosmic microwave background (WMAP) on the quartessence Chaplygin
model. We consider both adiabatic perturbations and intrinsic
non-adiabatic perturbations such that the effective sound speed
vanishes (\emph{Silent Chaplygin}). We show
that for the adiabatic case, only models with equation of state parameter $%
|\alpha |\lesssim 10^{-2}$ are allowed: this means that the allowed models
are very close to $\Lambda$CDM. In the \emph{Silent} case, however,
the results are consistent with observations in a much broader range, $%
-0.3<\alpha<0.7$.
\end{abstract}

\maketitle

\section{Introduction}

Two of the major puzzles of contemporary cosmology are the nature of dark
matter and of dark energy, the two big players in the cosmic arena. So far,
the only knowledge of these components refers to their density fraction and
to their equation of state, and even on these numbers we still have a large
uncertainty. Greater still is the ignorance of their clustering properties:
although we assume that essentially all the dark matter clusters in
observable objects and all the dark energy does remain quite homogeneous,
this is to a large extent only a simplification rather a consequence of
observations.

It is therefore no surprise that many works are currently devoted to the
possibility of merging the two puzzles into a single one: that is, finding a
single origin for both dark matter and dark energy. A possibility is to
assume an interaction between dark matter and dark energy \cite{ame} or to
fit both into a single complex field \cite{maibon}. However, these models
still contain two separate fields that account ultimately for the two
components. On a different level lies the hypothesis that there is a single
fluid that behaves as dark matter or dark energy according to the background
or the local density. Since there is only one unifying dark-matter-energy
component, besides baryons, photons and neutrinos, this model is usually
referred to as quartessence \cite{mak03a}. A phenomenological prototype of
quartessence models is the generalized Chaplygin model \cite{kam,bil,ben},
an exotic fluid with an inverse power law homogeneous equation of state
(EOS), $ \overline{p}=-M^{4(\alpha +1)}/\overline{\rho }^{\alpha }$, where $M
$ has dimension of mass and $\alpha >-1$ is a dimensionless parameter ($%
\alpha=1$ is the original Chaplygin gas). Such equation of state leads to a
component which behaves as dust in the past and as cosmological constant in
the future. For $\alpha=0$ the model reduces to $\Lambda$CDM \cite{ave03}.

For a wide range of values of the parameter $\alpha$, the
quartessence Chaplygin model is compatible
with several cosmological tests that involve only the background metric \cite%
{mak03b}. Nevertheless, problems may occur when one considers
perturbations. For instance, the CMB anisotropies spectrum is
strongly suppressed with respect to $\Lambda\mathrm{CDM}$ \cite{CF}.
Further, it was shown that, unless $\alpha$ is very much close to
$\alpha =0$, the mass power spectrum presents strong instabilities
and oscillations \cite{sand}. A quantitative CMB analysis
\cite{afbc} (see also \cite{BD} for a pre-WMAP analysis of the
generalized Chaplygin gas as dark energy) found that the parameter
$\alpha$ should be small ($\alpha <0.2$ at 95\% CL), even including
an additional CDM component, and smaller still without the latter.
We will show below that in fact this limit reduces to
$|\alpha|<0.01$. It is clear that this strong limit  is due to the
finite sound speed of the Chaplygin fluid: when it is sufficiently
large and positive, it prevents clustering on small scales and
therefore introduces a cut in the power spectrum which is at odds
with observations; when it is negative, on the other hand, the
clustering at small scales is enhanced beyond control.

It is important to stress, however, that unlike the background
tests, the perturbation analysis need further assumptions beyond the
equation of state. In \cite{reis03b}, it was shown that it is
possible to avoid the mass power spectrum problem if, for instance,
pressure perturbation vanishes. This can be done by introducing a
special type of intrinsic entropy perturbation such that the
effective sound speed \cite{hu} of the cosmic fluid vanishes (we
denote this as \emph{silent perturbations}). We call the Chaplygin
quartessence model with vanishing pressure perturbation \emph{Silent
Chaplygin} and refer to the standard case as\emph{\ adiabatic
Chaplygin}.

The main goal of this paper is to show that, unlike adiabatic Chaplygin,
Silent Chaplygin is consistent with CMB data for a wide range of parameters.
Besides CMB, we also consider consistency of Silent Chaplygin with current
data from large scale structure, from type Ia supernovae (SNIa) and baryon
fraction in galaxy clusters. Using the latest data sets, we review all these
tests using a careful treatment, presenting the outcome of a combined
analysis of the data for both Silent and adiabatic Chaplygin.

\section{Basic equations}

In this section we give a brief and basic description of the model. The
conservation equation for the generalized Chaplygin\ gas component in a
Robertson-Walker metric is solved by
\begin{equation}
\overline{\rho }_{Ch}=\overline{\rho }_{Ch0}[-w_{0}+\frac{1+w_{0}}{%
a^{3(\alpha +1)}}]^{\frac{1}{1+\alpha }},  \label{rhox}
\end{equation}%
where $a$ is the scale factor ($a=1$ today) and $w_{0}=-\frac{M^{4}}{%
\overline{\rho }_{Ch0}^{\alpha +1}}$ is the present equation of state.

The equation-of-state parameter ($w=\overline{p}/\overline{\rho }$) and the
adiabatic sound velocity ($c_{s}^{2}=d\overline{p}/d\overline{\rho }$) for
the Chaplygin\ gas, are given by
\begin{equation}
w_{Ch}(a)=\frac{w_{0}}{-w_{0}+(1+w_{0}){a^{-3(\alpha +1)}}}  \label{wx}
\end{equation}%
and
\begin{equation}
c_{sCh}^{2}(a)=-\alpha \,w_{Ch}(a).
\end{equation}

From the equation above, it is clear that at early times, when
$a\rightarrow 0,$ we have $w_{Ch}\rightarrow 0$, and the fluid
behaves as non relativistic matter. At late times, when $a\gg 1$, we
obtain $w_{Ch}\rightarrow -1$. Further, the adiabatic sound speed
has maximum value when the equation of state parameter is minimum.
Therefore, the epoch of cosmic acceleration, in
these models, coincides with that of large $c_{s}^{2}$. As remarked in \cite%
{reis05}, this is a characteristic of all unified models in which the
equation of state is convex, i.e., is such that $\frac{d^{2}\overline{p}}{d%
\overline{\rho }^{2}}=\frac{dc_{s}^{2}}{d\overline{\rho }}<0$. However, this
property is not mandatory for a general fluid. Models with concavity
changing equations of state, may have $c_{s}^{2}$ negligibly small when the
energy density reaches its minimum value. This is an important property that
must be considered in constructing acceptable adiabatic quartessence models
\cite{makler05}.

The perturbation equations for a fluid with equation of state $w$ and
adiabatic sound speed $c_{s}^{2}$ in \emph{synchronous} gauge are (for the
sake of simplicity, here we neglect baryons and radiation, and we assume
that both spatial curvature and the anisotropic perturbation vanish)
\begin{equation}
\delta ^{\prime }=-3(w\Gamma +c_{s}^{2}\delta -w\delta )-(1+w)(\theta +\frac{%
h_{L}^{\prime }}{2})
\end{equation}%
\begin{equation}
\theta ^{\prime }=\frac{k^{2}}{(1+w)\mathcal{H}^{2}}(w\Gamma
+c_{s}^{2}\delta )+(3c_{s}^{2}-1-\frac{\mathcal{H}^{\prime }}{\mathcal{H}}%
)\theta
\end{equation}%
\begin{equation}
h_{L}^{\prime }=\frac{2k^{2}\eta }{^{\mathcal{H}^{2}}}+3\delta \Omega
\end{equation}%
\begin{equation}
\eta ^{\prime }=\frac{3(1+w)\mathcal{H}^{2}\theta \Omega }{2k^{2}}
\end{equation}%
where derivatives are with respect to $\log a$. Here $\delta $ is
the density contrast, $\Omega $ is the density parameter, $h_{L}$
and $\eta $ are metric perturbations, $\theta $ is the divergence of
the velocity perturbation, $\Gamma$ is the entropy perturbation and
$\mathcal{H}\equiv da/d\tau $, $\tau $ being the conformal time. To
these equations, we must add the equations for baryons and
relativistic fluid. The assumption of a \emph{silent universe}
requires that \cite{reis03b}
\begin{equation*}
\delta p=\overline{p}\Gamma +c_{s}^{2}(\delta \rho )=0
\end{equation*}%
so that
\begin{equation*}
\Gamma =-\frac{c_{s}^{2}}{w}\delta
\end{equation*}%
We adopt therefore these equations with $w=w_{Ch}$ and $c_{s}=c_{sCh}$. With
this choice, $\Gamma =\alpha \delta $ and models with $\alpha =0$ are both,
silent and adiabatic.

\section{Observational constraints}

In the subsections below, we will derive constraints on the parameters $%
\alpha $ and $w_{0}$ from four data sets: SNIa, X-ray cluster gas fraction,
SDSS power spectrum and CMB.

\subsection{SNIa}

The luminosity distance of a light source is defined in such a way as to
generalize to an expanding and curved space the inverse-square law of
brightness valid in a static Euclidean space,
\begin{equation}
d_{L}=\left(\frac{L}{4\pi \mathcal{F}}\right)^{1/2}.  \label{dL}
\end{equation}
In Eq. (\ref{dL}), ${L}$ is the absolute luminosity and $\mathcal{F}$ is the
measured flux.

For a source of absolute magnitude $M$, the apparent bolometric magnitude $%
m(z)$ can be expressed as
\begin{align}
m(z)=& M-5\log H_{0}+25+5\log D_{L}  \notag \\
=& M+42.3841-5\log h\,+5\log D_{L},  \label{appmag}
\end{align}%
where $D_{L}$ is the luminosity distance in units of $H_{0}^{-1}$(we are
using $c=1$),
\begin{equation}
D_{L}=\int_{0}^{z}\left[ E\left( z^{\prime }\right) \right] ^{-1}dz^{\prime
}\,,  \label{DL}
\end{equation}%
where
\begin{align}
& E^{2}\left( z\right) =\left\{ \omega _{b}h^{-2}\left( 1+z\right)
^{3}\right.  \notag \\
& +\left. (1-\omega _{b}h^{-2})\left[ -w_{0}+(1+w_{0})\left( 1+z\right)
^{3\left( 1+\alpha \right) }\right] ^{\frac{1}{(1+\alpha )}}\right\} .
\label{EzTot}
\end{align}%
Here, $z$ is the redshift, $\omega _{b}=\Omega _{b}h^{2}$, and $h$ is the
Hubble constant in units of $100km/sMpc^{-1}$.

\begin{figure}[tbp]
\centering
\par
\hspace*{+0.3in}
\par
\includegraphics[height= 7.0
cm,width=8.0cm]{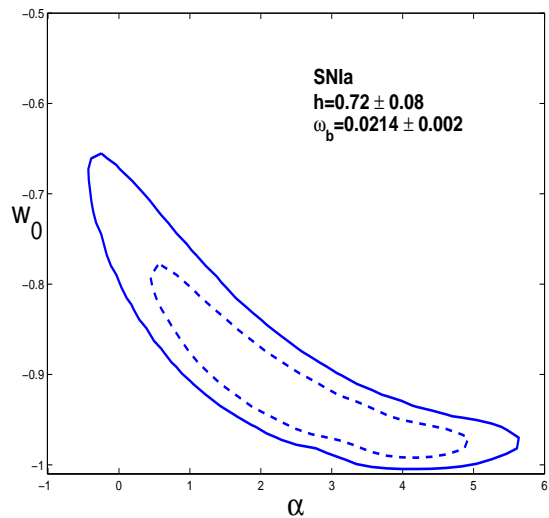}
\caption{Constant confidence contours ($68\%$ and $95\%$) in the ($\protect%
\alpha ,w_{0}$) plane allowed by SNIa data, as described in the text}
\label{sne}
\end{figure}

For the SNIa analysis, we use the `gold' data set of Riess \emph{et al.}
\cite{riess04}. The data in this sample is given in terms of the extinction
corrected distance modulus, $\mu _{0}=m-M$. To obtain the likelihood of the
parameters, we use a $\chi ^{2}$ statistics such that,
\begin{align}
& \chi ^{2}(\alpha ,w_{0},h,\omega _{b})=  \notag \\
& \sum\limits_{i=1}^{157}\left( \frac{\mu _{0,i}-42.3841+5\log h\,-5\log
D_{L}(z_{i,}\alpha ,w_{0},h,\omega _{b})}{\sigma _{\mu _{0,i}}}\right) ^{2}.
\label{chi2}
\end{align}%
In (\ref{chi2}), $\sigma _{\mu _{0,i}}$ are the estimated errors in the
individual distance moduli, including uncertainties in galaxy redshift and
also taking into account the dispersion in supernovae redshift due to
peculiar velocities (see \cite{riess04} for details). To determine the
likelihood of the parameters $\alpha $ and $w_{0}$, we marginalize the
likelihood function over $h$ and $\omega _{b}$. We use two different priors
in this work. We consider flat priors when combining the SNIa and clusters
results with CMB, while in this and in the next subsection we adopt Gaussian
priors such that $h=0.72\pm 0.08$ \cite{hst} and $\omega _{b}=0.0214\pm
0.002 $ \cite{kir03}. The results of our SNIa analysis are displayed in Fig. %
\ref{sne}. In the figure, we show $68\%$ and $95\%$ confidence level
contours in the $(\alpha ,$ $w_{0})$ plane.

\subsection{X-ray Cluster Gas Fraction}

Clusters of galaxies are the most recent large-scale structures formed and
the largest gravitationally bound systems known. Therefore, the
determination of their matter contents is important since cluster properties
should approach those of the Universe as a whole. By measuring the baryon
mass fraction $\Omega _{b}/\Omega _{m}$ in rich clusters, and combining this
ratio with $\Omega _{b}$ determinations from primordial nucleosynthesis,
constraints on $\Omega _{m}$ can be placed \cite{white}. Further, by
assuming that baryon mass fraction in clusters of galaxies is independent of
redshift, it is also possible to constrain the geometry and, consequently,
the dark energy density. A method based on this idea was suggested by Sasaki
\cite{sasaki} and Pen \cite{pen} and further developed and applied by Allen,
Schmidt, and Fabian \cite{allen02}.

In this section we use the new data set of Allen \emph{et al.}
\cite{allen04} to constraint the Chaplygin models. These authors
extracted from Chandra observations the x-ray gas mass fraction
($f_{\text {gas}}$) of twenty six
massive, dynamically relaxed galaxy clusters, with redshifts in the range $%
0.08<z<0.9$, and that have converging $f_{\text {gas}}$ within a radius $%
r_{2500}$ (radius encompassing a region with mean mass density $2500$ times
the critical density of the Universe at the cluster redshift).

To determine the confidence region of the parameters of the model, we use
the following $\chi ^{2}$ function in our computation:
\begin{equation}
\chi ^{2}(\alpha ,w_{0},h,\omega _{b},b)=\sum \limits _{i=1}^{26}\frac{\left[%
f_{\text {gas}}^{\text {mod}}(z_{i},\alpha ,w_{0},h,\omega _{b},b)-f_{\text {%
gas},i}\right]^{2}}{\sigma _{f_{\text {gas},i}}^{2}}\, ,  \label{chi2cluster}
\end{equation}
where $z_{i}$, $f_{\text {gas},i}$, and $\sigma _{f_{\text {gas},i}}$ are,
respectively, the redshifts, the SCDM ($h=0.5$) best-fitting values, and the
symmetric root-mean-square errors for the 26 clusters as given in \cite%
{allen04}. In Eq.~(\ref{chi2cluster}), $f_{\text {gas}}^{\text {mod}}$ is
the model function \cite{allen02}%
\begin{equation}
f_{\text {gas}}^{\text {mod}}=\frac{b\omega _{b}h^{-2}}{(1+0.19\sqrt{h}%
)\Omega _{m}^{\text {eff}}}\left(\frac{h}{0.5}\frac{d_{A}^{\text {EdS}}}{%
d_{A}^{\alpha ,A}}\right)^{3/2}\, .  \label{fgasmod}
\end{equation}
Here, $d_{A}=(1+z)^{-2}d_{L}$ is the angular diameter distance to the
cluster, $b$ is a bias factor that takes into account the fact that the
baryon fraction in clusters could be lower than for the Universe as a whole
and
\begin{equation}
\Omega _{m}^{\text {eff}}=(1-\omega
_{b}h^{-2})\left(1+w_{0}\right)^{1/\left(1+\alpha \right)}+\omega
_{b}h^{-2}\, .  \label{OmMeff}
\end{equation}
is the effective matter density parameter \cite{mak03b}. In our
computations, we first marginalize analytically over the bias factor
assuming that it is Gaussian-distributed with $b=0.824\pm 0.089$ \cite%
{allen04,rapetti}. To determine the likelihood of the parameters $\alpha $
and $w_{0}$, we next marginalize the likelihood function over $h$ and $%
\omega _{b}$. As remarked before, we assume here a Gaussian prior such that $%
h=0.72\pm 0.08$ \cite{hst} and $\omega _{b}=0.0214\pm 0.002$ \cite{kir03}.

\begin{figure}[tbp]
\centering
\par
\hspace*{+0.3in}\includegraphics[  width=8cm,
  height=7cm]{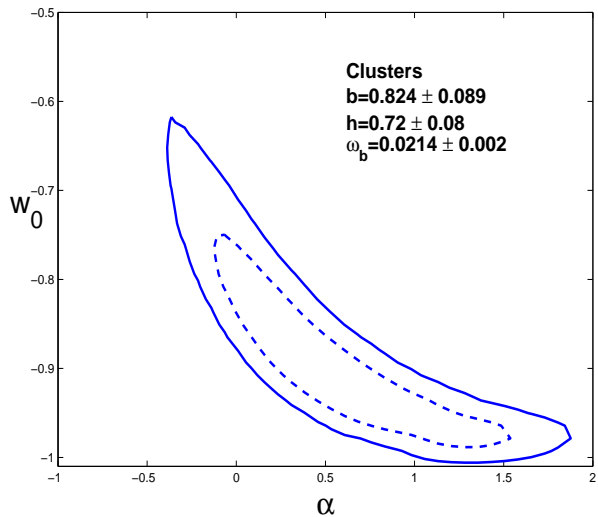}
\caption{Constant confidence contours ($68\%$ and $95\%$) in the ($\protect%
\alpha ,w_{0}$) plane from cluster $f_{gas}$ data as described in the text.}
\label{clusters}
\end{figure}

In Fig. \ref{clusters}, we show the $68\%$ and $95\%$ confidence contours on
the parameters $\alpha $ and $w_{0}$ determined from the Chandra data.

\subsection{CMB}

\begin{figure}[tbp]
\centering
\par
\hspace*{+0.3in}%
\includegraphics[  width=8cm,
  height=7cm]{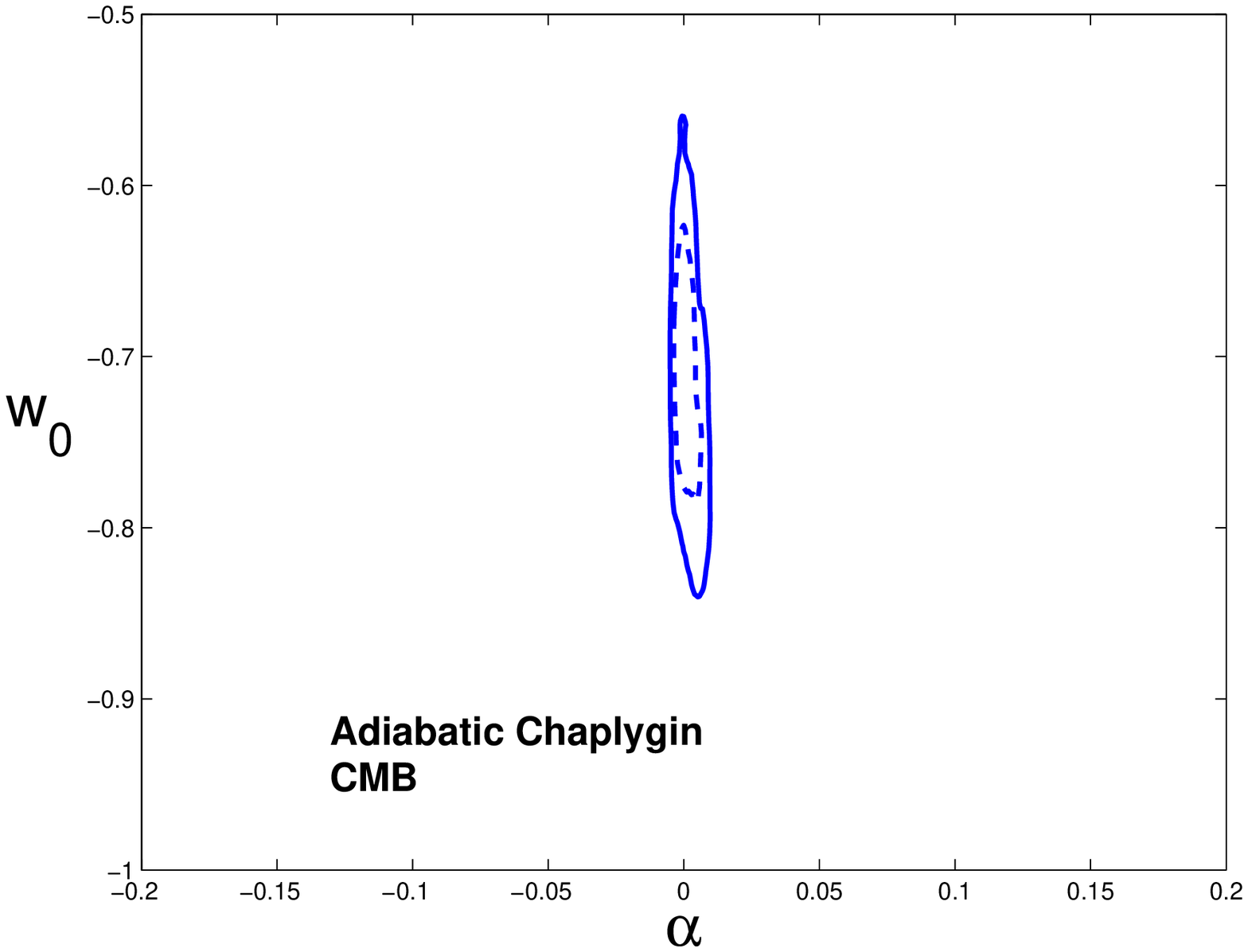}
\caption{Constant confidence contours ($68\%$ and $95\%$) in the ($\protect%
\alpha ,w_{0}$) plane allowed by CMB, as described in the text.}
\label{adiabaticcmb}
\end{figure}

\begin{figure}[tbp]
\centering
\par
\hspace*{+0.3in}%
\includegraphics[    width=5cm,
  height=5cm]{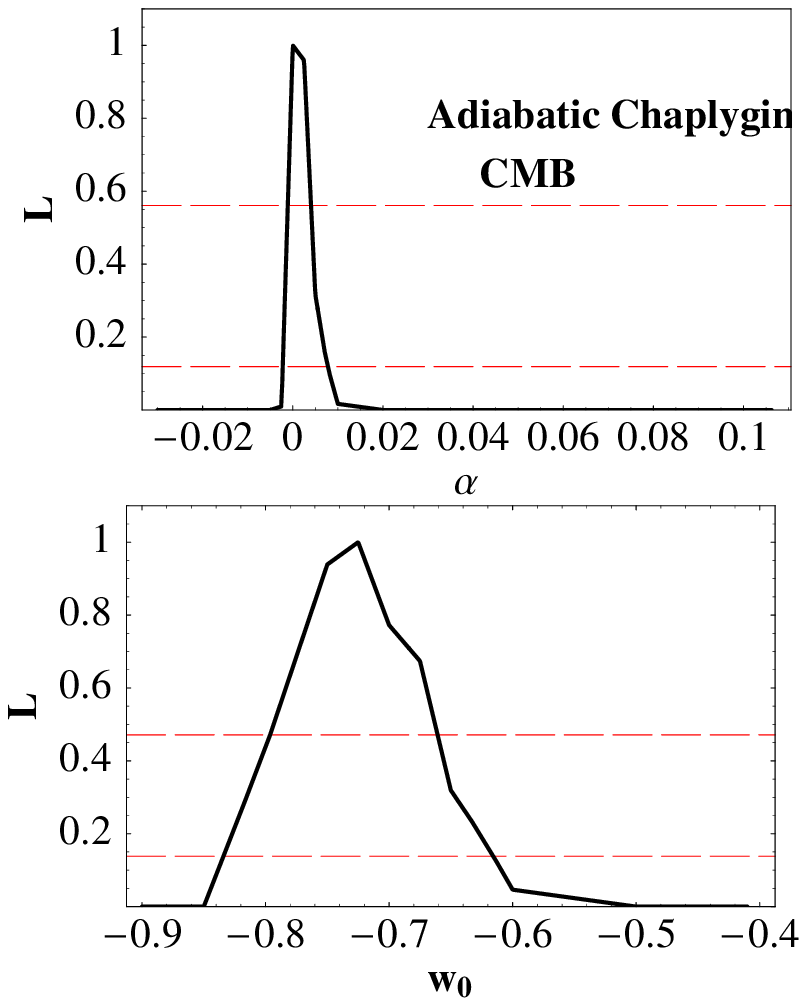}
\caption{CMB likelihood functions for $\protect\alpha $ and $w_{0}$ in the
adiabatic case. The horizontal lines mark the 68$\%$ and 95$\%$ c.l., top to
bottom.}
\label{adiab1dlik}
\end{figure}

\begin{figure}[tbp]
\centering
\par
\hspace*{+0.3in}\includegraphics[  width=8cm,
  height=7cm]{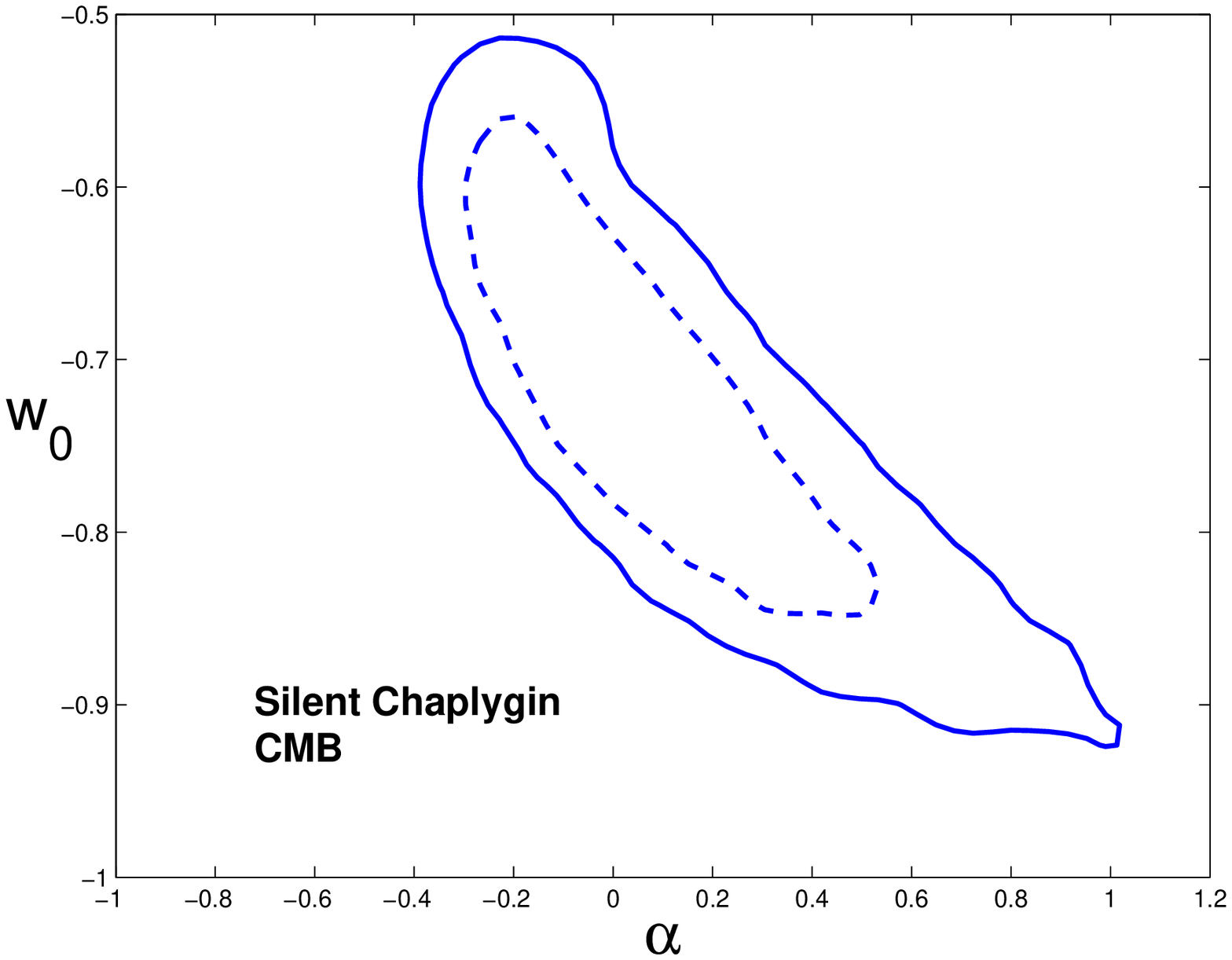}
\caption{Constant confidence contours ($68\%$ and $95\%$) in the ($\protect%
\alpha ,w_{0}$) plane allowed by CMB, as described in the text}
\label{silentcmb}
\end{figure}

Here we compare the model to the combined temperature and polarization power
spectrum estimated by WMAP \cite{hin}. To derive the likelihood we adopt a
version of the routine described in Verde \emph{et al.} \cite{verde}, which
takes into account all the relevant experimental properties (calibration,
beam uncertainties, window functions, etc).

Our theoretical model depends on two Chaplygin parameters, four cosmological
parameters and the overall normalization $N$:
\begin{equation}
\alpha ,w_{0},n_{s},h,\omega _{b},\tau ,N.
\end{equation}
The overall normalization has been integrated out numerically. We calculate
the theoretical $C_{\ell ,t}$ spectra by a modified parallelized CMBFAST
\cite{sel} code that includes the full set of perturbation equations
\cite{CF,afbc} with the addition of non-adiabatic pressure perturbations.
We do not include gravitational waves and the other parameters are set as
follows: $T_{cmb}=2.726K,$ $Y_{He}=0.24,N_{\nu }=3.04$.

We evaluated the likelihood on a unequally spaced grid of roughly $50,000$
models (for each normalization) with the following top-hat broad priors: $%
w_{0}\in (-1,-0.5),$ $\quad \alpha \in (-0.8,1.5), \quad $\ $n_{s}\in
(0.8,1.2),$ $\quad \omega _{b}\in (0.02,0.025),\quad $\ $\tau \in (0.,0.3)$.
For the Hubble constant we adopted the top-hat prior $h\in (0.6,0.8);$ we
also employed the HST result \cite{hst} $h=0.72\pm 0.08$ (Gaussian prior).

In Fig. \ref{adiabaticcmb}, we show the confidence region on $\alpha ,w_{0}$
in the adiabatic case, after marginalization over all the other parameters
with flat priors: as it can be seen, this is the most stringent test among
those studied in this paper. As anticipated, it restricts $\alpha$ to be
very close to 0: in other words, observations of CMB demands that the
background of adiabatic Chaplygin be almost indistinguishable from $\Lambda$%
CDM. In Fig. \ref{silentcmb} we contrast this result with the silent
case: now the allowed region for $\alpha $ widens a lot,
encompassing values close to unity. What is particularly interesting
is that even with $\alpha =1$, Silent Chaplygin is consistent at the
2$\sigma $ level, while it was
obviously ruled out in the adiabatic case. In Figs. \ref{adiab1dlik} and \ref%
{silent1dlik}. we show the one-dimensional likelihood for $\alpha $ and $%
w_{0}$, with marginalization over all the other parameters.

\begin{figure}[tbp]
\centering
\par
\hspace*{+0.3in}%
\includegraphics[   width=5cm,
  height=5cm]{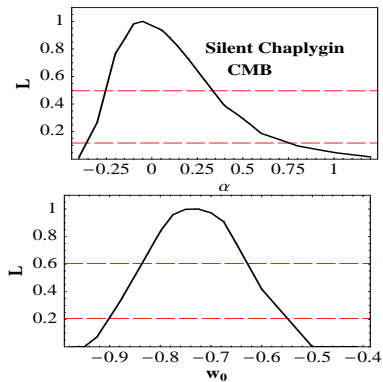}
\caption{CMB likelihood functions for $\protect\alpha $ and $w_{0}$ in the
silent case. The horizontal lines mark the 68$\%$ and 95$\%$ c.l., top to
bottom.}
\label{silent1dlik}
\end{figure}

\subsection{Large scale structure}

\begin{figure}[tbp]
\centering
\par
\hspace*{+0.3in}
\par
\includegraphics[ width=7cm,
  height=6cm
]{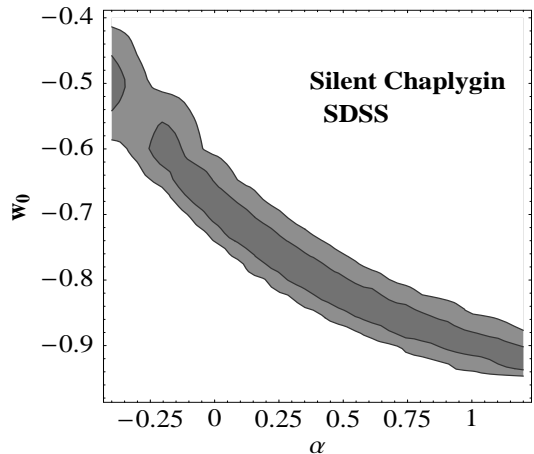}
\caption{SDSS likelihood functions at 68$\%$ and 95 $\%$ c.l. for $\protect%
\alpha $ and $w_{0}$ in the silent case.}
\label{silent1sdss}
\end{figure}

The results of Sandvik et al. (2004) \cite{sand} have shown that the matter
power spectrum of adiabatic Chaplygin is plagued by strong instabilities and
oscillations for any $\alpha $ significantly different from zero, leading to
a stringent upper limit to $|\alpha |$. In \cite{reis03b}, it has been shown
that these problems do not occur for the silent model. To test this in a
more quantitative way we perform here a simplified likelihood analysis with
the following procedure. We compared the baryon spectra of the silent case
to the matter power spectrum of Sloan Digital Sky Survey (SDSS) as obtained
in Tegmark et al. (2004) \cite{teg04}, cutting at $k=0.02h/$Mpc, using the
likelihood routine provided by M. Tegmark \cite{tegweb}, and marginalizing
over the amplitude (i.e., we are comparing only the slope of the spectrum,
not its absolute value). In order to save computer time we restricted
ourselves to a subset of parameters, fixing $\omega _{b}=0.023$, $n_{s}=1$
and $h=0.7$. As we show in Fig. \ref{silent1sdss}, the results are rather
similar to the CMB case. Since we used only a very small subset of the whole
parameters space we will not include the SDSS likelihood in the combined
likelihood of the next section. The loss in constraining power is acceptable
since one can easily see that the CMB and the SDSS confidence regions are
overlapping to a good extent. Moreover, extending the range of parameters as
in the CMB case the SDSS constraints would become weaker and would not add
much to the information from the CMB.

\begin{figure}[tbp]
\centering
\par
\hspace*{+0.3in}
\par
\includegraphics[width=7cm,
  height=6cm
]{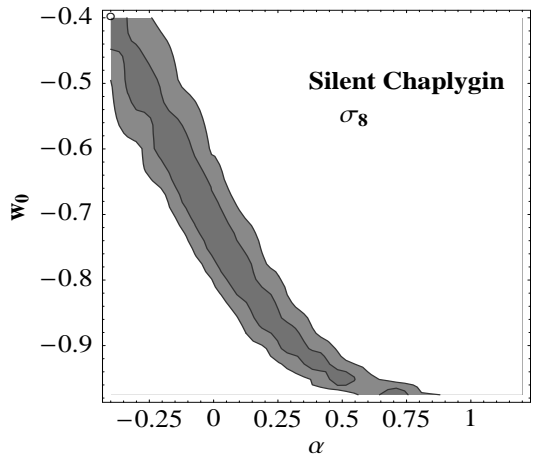}
\caption{Likelihood function for $\protect\sigma_8$ at 68$\%$ and 95 $\%$
c.l. for $\protect\alpha $ and $w_{0}$ in the silent case.}
\label{s8}
\end{figure}

We also calculated the baryon fluctuation variance $\sigma_8$ (i.e.
the absolute normalization of the spectrum) for the same subset of
parameters. In Fig. \ref{s8} we show the contour plots of the
likelihood
assuming a Gaussian distribution of the observed $\sigma_8$ with $%
\sigma_8=0.9\pm0.1$. This test is meant to be only qualitative since the
value of $\sigma_8$ that is derived from data is strongly degenerate with $%
\Omega_m$ and with the bias factor and/or it depends on calibration obtained
with $N$-body simulations based on some specific cosmological model. We only
notice that for the Silent Chaplygin case there exists a large region in our
parameter space which accounts for values of $\sigma_8$ which are consistent
with the standard estimation. It is nevertheless interesting to note that
the $\sigma_8$ analysis shows preference for lower values of $w_0$ (for a
fixed $\alpha$) than the CMB test. This could help to further reduce the
parameters space.

\section{Combined Constraints and Conclusion}

\begin{figure}[tbp]
\centering
\par
\hspace*{+0.3in}
\par
\includegraphics[height= 7.0
cm,width=8.0cm]{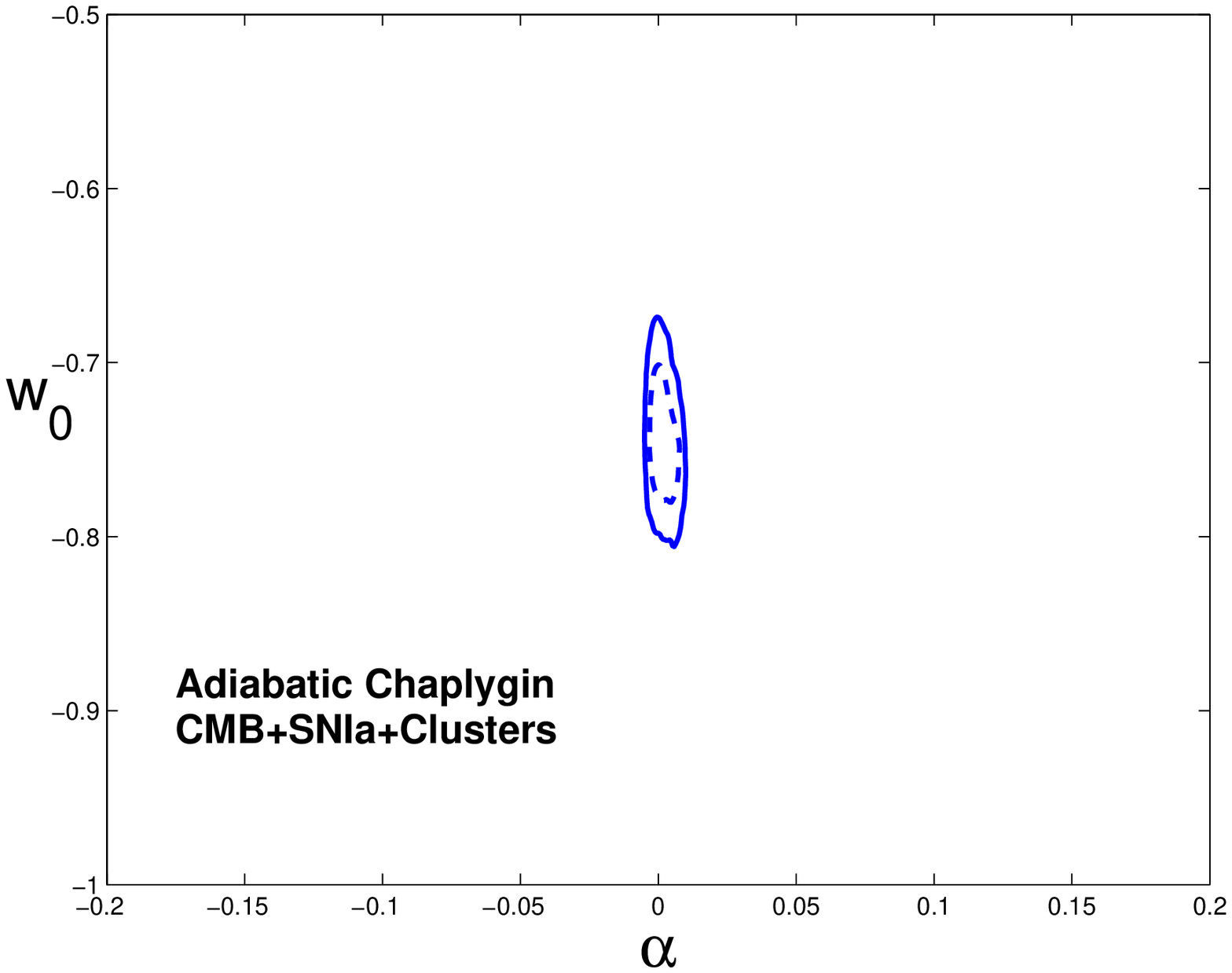}
\caption{Constant confidence contours ($68\%$ and $95\%$) in the ($ \protect%
\alpha ,w_{0}$) plane allowed by SNIa, clusters and CMB, as described in the
text.}
\label{combadiabatic}
\end{figure}

\begin{figure}[tbp]
\centering
\par
\hspace*{+0.3in}
\par
\includegraphics[height= 7.0
cm,width=8.0cm]{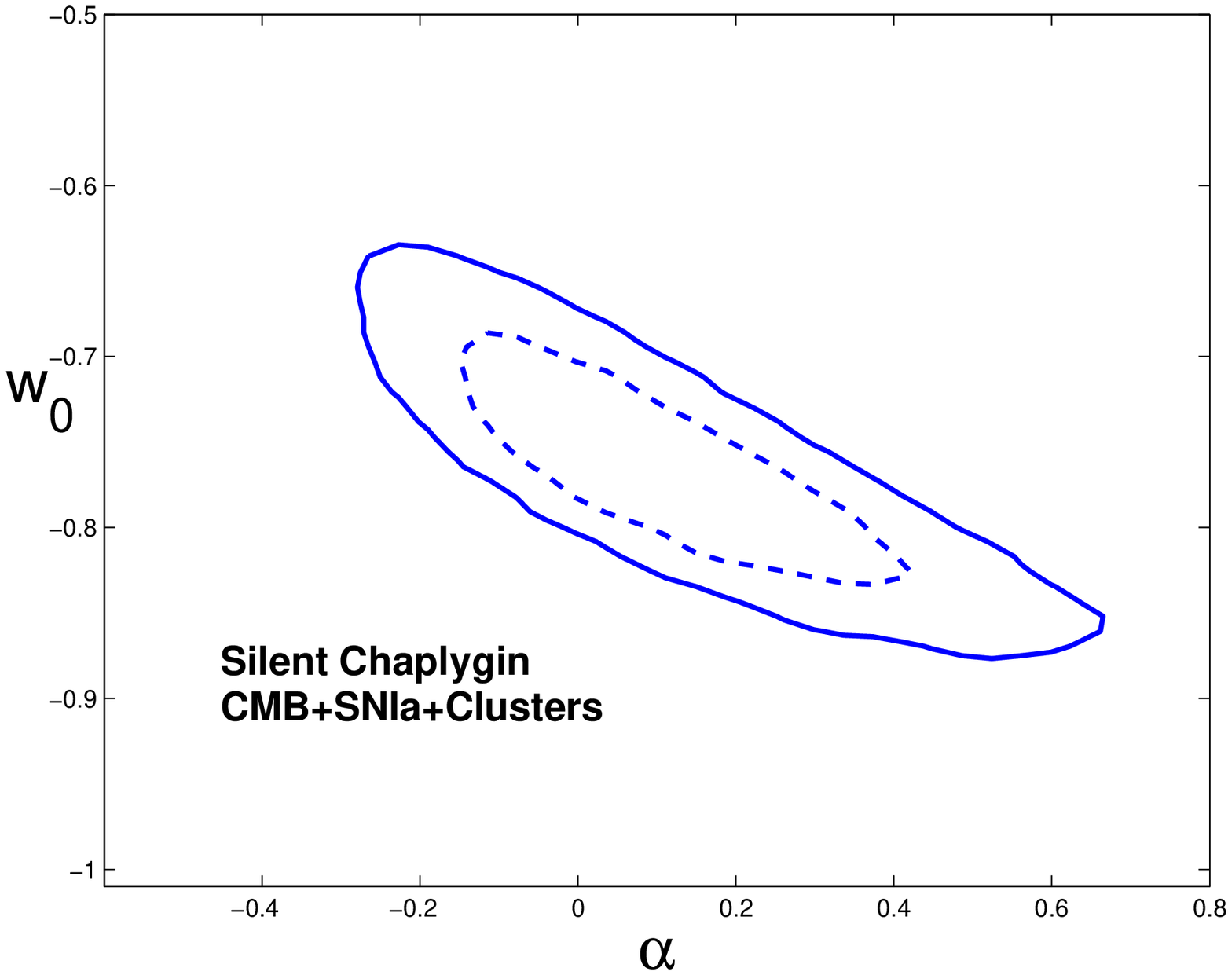}
\caption{Constant confidence contours ($68\%$ and $95\%$) in the ($ \protect%
\alpha ,w_{0}$) plane allowed by SNIa, clusters and CMB, as described in the
text}
\label{combsilent}
\end{figure}

\begin{figure}[tbp]
\centering
\par
\hspace*{+0.3in}
\par
\includegraphics[ width=5cm,
  height=5cm]{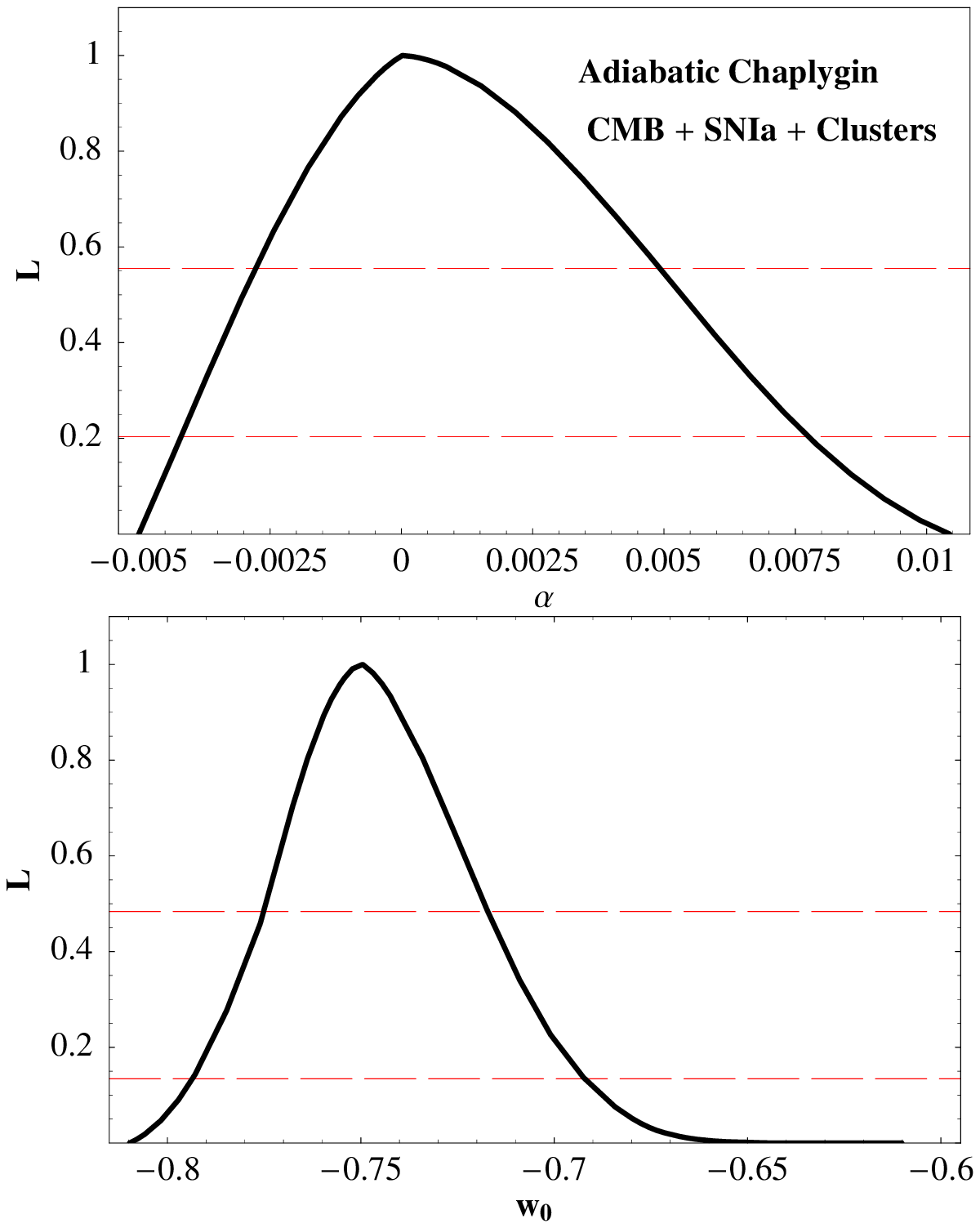}
\caption{Likelihood functions for $\protect\alpha $ and $w_{0}$ in the
adiabatic case for SNIa, clusters and CMB. The horizontal lines mark the 68$%
\%$ and 95$\%$ c.l., top to bottom.}
\label{1dimadiabatic}
\end{figure}

\begin{figure}[tbp]
\centering
\par
\hspace*{+0.3in}
\par
\includegraphics[ width=5cm,
  height=5cm]{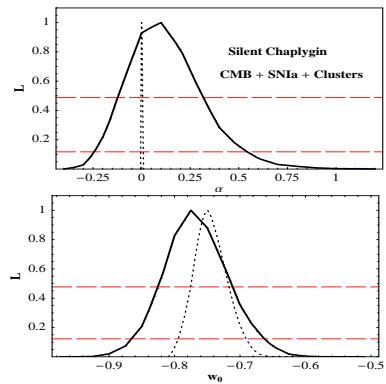}
\caption{Likelihood functions for $\protect\alpha $ and $w_{0}$ in
the silent case for SNIa, clusters and CMB. The horizontal lines
mark the 68$\%$ and 95$\%$ c.l., top to bottom. For comparison, we
plot as dotted lines the combined likelihood functions  of the
adiabatic case.} \label{1dimsilent}
\end{figure}

Here we present a combined analysis of the constraints discussed in the
previous section. In Fig. \ref{combadiabatic} we display the allowed region
of the parameters $w_{0}$ and $\alpha $ from a combination of data from
SNIa, clusters and CMB in adiabatic Chaplygin. The constraints on the
parameter $\alpha $ are roughly the same as those obtained from CMB alone.
Only models with $\alpha $ very close to the $\Lambda $CDM limit are
allowed. Although including SNIa and clusters in the CMB analysis almost do
not affect the constraints on $\alpha $, they tighten the constraints on the
parameter $w_{0}$, reducing even more the allowed region of the parameters
space in the adiabatic case. The final result (95\% c.l.) for the adiabatic
case is
\begin{equation}
-0.005<\alpha<.01,\quad -0.8<w_0<-0.7.
\end{equation}
In contrast, in Fig. \ref{combsilent}, the Silent Chaplygin model is
consistent with the observables considered here for a wide range of
parameters. The constraints in the silent case are (95\% c.l.)
\begin{equation}
-0.25<\alpha<0.5,\quad -0.87<w_0<-0.68,
\end{equation}
with a remarkable fifty-fold extension in the range of $\alpha$. The
one-dimensional likelihood functions are shown in Figs. \ref{1dimadiabatic}
and \ref{1dimsilent}.

The idea of unifying dark matter and dark energy through a single component
has motivated many works in the last few years. Most of these investigations
concentrated their efforts in analyzing the generalized Chaplygin fluid,
that is considered the prototype of the quartessence models. After the
results of \cite{CF,sand,afbc}, it became clear that adiabatic Chaplygin,
as
quartessence, is ruled out unless the parameter $\alpha$ is very close to
zero. In the present paper we confirmed this result by using current CMB
data, extending the quantitative results of \cite{afbc} to quartessence
models. In fact, for an effective sound speed different from zero, any
quartessence model with a convex EOS will suffer the same kind of problem
\cite{reis05}. There are two simple ways to get around this problem:
choosing an EOS not convex \cite{makler05} or introducing entropic
perturbations as we did in this paper. This shows that indeed quartesence
models of the Chaplygin type (one fluid, two parameters) can be considered
as real alternatives to $\Lambda$CDM (two fluids, one parameter). It is
clear however that these phenomenological models should be further
investigated: the challenge is to connect them to a more fundamental theory.

Finally, we remark that one possible source of problems for Silent Chaplygin
is lensing skewness \cite{reis04}. It would be remarkable if higher-order or
non-linear effects would prove necessary to rule out, or strongly constrain,
these models.

\begin{acknowledgments}
We thank Maur\'{i}cio Calv\~ao, Roberto Colistete, Martin Makler and
Ribamar Reis for useful discussions. The CMB computations have been
performed at CINECA (Italy) under the agreement INAF@CINECA. We
thank the staff for support. IW is partially supported by the
Brazilian research agency CNPq.
\end{acknowledgments}

\end{document}